\documentclass[useAMS,usenatbib]{mn2e}
\usepackage{graphics,rotating,multirow,bm,color,hyperref,amsmath}
\usepackage{subfig}
\hypersetup{
    colorlinks=true,
    linkcolor=black,
    citecolor=black,
}

\title[Excursion set with correlated steps for modified gravity]
  {Excursion set theory for modified gravity: correlated steps,  mass functions and halo bias}
\author[T. Y. Lam \& B. Li]
  {Tsz Yan Lam$^1$\thanks{E-mail: tszyan.lam@ipmu.jp}, Baojiu~Li$^{2,1}$\thanks{E-mail: baojiu.li@durham.ac.uk}\\
  $^1$Kavli-IPMU, University of Tokyo, Kashiwa, Chiba 277-8583, Japan\\
  $^{2}$Institute for Computational Cosmology, Department of Physics, Durham University, South Road, Durham DH1 3LE, UK}
  
\pagerange{\pageref{firstpage}--\pageref{lastpage}} \pubyear{2012}

\def\LaTeX{L\kern-.36em\raise.3ex\hbox{a}\kern-.15em
    T\kern-.1667em\lower.7ex\hbox{E}\kern-.125emX}

\newcommand{\tcr}[1]{\textcolor{black}{#1}}

\begin{document}

\label{firstpage}

\maketitle

\begin{abstract}
We show how correlated steps 
introduces significant contributions to the modification of 
the halo mass function in modified gravity models, taking 
the chameleon models as an example, in the framework of 
the excursion set approach.
This correction applies to both Lagrangian and Eulerian 
environments discussed in previous studies.
Correlated steps also enhances the modifications 
due to the fifth force 
in the conditional mass function as well as the halo bias.
We found that abundance and clustering measurements from different
environments can provide strong constraints on 
the chameleon models.
\end{abstract}


\section{Introduction}

\label{sect:intro}
The discovery of accelerated expansion of the Universe 
sparks surge of research on the possibility of modified
gravity model \citep[see, for example,][for review]{jk10,cliftonetal}.
The main goal of such modification is to alter the large
scale behaviour to explain the weakening of gravity -- 
however any modifications in the gravity model must at the same time 
satisfy the 
tight constraints from the solar system test.
One way to fullfill this requirement is to include some 
kinds of screening mechanisms to suppress the modification in 
local environment (high density regime).
In this work we focus on one particular example -- one that modifies
gravity by introducing 
a dynamical scale field which mediates a fifth force. 
This scalar field has the nice property such that the fifth force 
is strongly suppressed in regions of high matter density and hence
it passes the solar system test.

The chameleon model of \citet{kw,ms} is a representative example (other noticeable examples are the environmentally dependent dilation \cite{bbds2010} and symmetron \citep{hk2010} models). 
The background evolution and the linear perturbations on large scale 
can be indistinguishable from the 
standard $\Lambda$CDM cosmology \citep{hs2007,lb2007,bbds2008,lz2009};
solar system test is satisfied by construction. 
Hence nonlinear structure formation is the only regime where
effects of such models would possibly be detected.
A number of studies 
\citep{oyaizu2008,olh2008,sloh2009,lz2009,lz2010,lb2011,zlk2011,lztk2011,bbdls2011,dlmw2011}
employed N-body numerical simulations to study nonlinear structure formation
-- however high resolution simulations with cosmological volume 
are still challenging due to the non linear equation governing the 
scalar field.

This work aims at investigating the effect of chameleon mechanism on 
large scale structure. 
We apply the excursion set approach \citep{bcek,mw1996,st1999}
to compute the halo mass function in the same line as previous
analyses done by \citet{le2012,ll2012}.
\citet{le2012} first illustrated the idea of extending the 
excursion set approach to calculate halo mass function in chameleon models (see \citet{brs2010} for an earlier work in this direction, where the authors studied the spherical collapse in chameleon models in detail). 
They did so by assuming a fixed 
Lagrangian environment. \citet{ll2012} improved this 
approach by introducing an Eulerian environment. This choice avoids
the unphysical  requirement that all environments having the same mass
(which also places an upper limit of the halo mass). 
The calculation involves two first crossing distributions, 
one for the Eulerian 
environment and the other one for the (modified) halo formation
barrier. 
Both studies compute the effect of the chameleon mechanism
assuming uncorrrelated steps 
in the excursion set approach 
-- it is the main focus of the first half of 
the current 
work to investigate how correlated steps in the excursion set approach
interact with the fifth force and the chameleon mechanism. 
Recently there are renewed interests 
in going beyond the uncorrelated steps excursion 
set in GR+$\Lambda$CDM cosmology 
to compute the halo/void abundance \citep{mr10,pls2012a,pls2012b,ms12} 
and the halo bias \citep{maetal11,ps2012,ms12}.

In addition to the unconditional mass function, 
the excursion set approach lays down the framework to calculate 
 the conditional mass function as well as the halo bias. 
The latter is of particular importance since it relates
observables (distribution of halos) to the underlying matter
distribution.
Since the mass function in chameleon models depends on 
the environment density, one may expect
the conditional mass function (and hence the halo bias) to be modified. 
The second part of this paper discusses how the conditional 
mass function as well as the halo bias depend on the chameleon model.
 
This paper is organised as follows: in \S~\ref{sect:chameleon} 
we briefly review the theoretical model to be considered and summarise
its main ingredients.
The effect of the chameleon mechanism on the halo mass function 
within the framework of excursion set approach is presented in 
\S~\ref{sect:correlated}, where we also include analytical approximation 
for correlated steps in \S~\ref{sect:analytical}.
\S~\ref{sect:condition} discusses the conditional mass function and halo bias
in chameleon models where we show measurements in underdense environment
provide a good test for chameleon signatures.
Finally we conclude in \S~\ref{sect:con}.

\section{The Chameleon Theory}

\label{sect:chameleon}

This section lays down the theoretical framework for investigating the effects of coupled scalar field(s) in cosmology.  We shall present the relevant general field equations in \S~\ref{subsect:equations}, and then
specify the models analysed in this paper  in \S~\ref{subsect:specification}.

\subsection{Cosmology with a Coupled Scalar Field}

\label{subsect:equations}

The equations presented in this sub-section can be found in \citep{lz2009, lz2010, lb2011}, and  are presented here only to make this work self-contained.

We start from a Lagrangian density
\begin{eqnarray}\label{eq:lagrangian}
{\cal L} =
{1\over{2}}\left[{R\over\kappa}-\nabla^{a}\varphi\nabla_{a}\varphi\right]
+V(\varphi) - C(\varphi){\cal L}_{\rm{DM}} +
{\cal L}_{\rm{S}},
\end{eqnarray}
in which $R$ is the Ricci scalar, $\kappa=8\pi G$ with $G$ being the
gravitational constant, ${\cal L}_{\rm{DM}}$ and
${\cal L}_{\rm{S}}$ are respectively the Lagrangian
densities for dark matter and standard model fields. $\varphi$ is
the scalar field and $V(\varphi)$ its potential; the coupling
function $C(\varphi)$ characterises the coupling between $\varphi$
and dark matter. Given the functional forms for $V(\varphi)$ and $C(\varphi)$
a coupled scalar field model is then fully specified.

Varying the total action with respect to the metric $g_{ab}$, we
obtain the following expression for the total energy momentum
tensor in this model:
\begin{eqnarray}\label{eq:emt}
T_{ab} &=& \nabla_a\varphi\nabla_b\varphi -
g_{ab}\left[{1\over2}\nabla^{c}\nabla_{c}\varphi-V(\varphi)\right]\nonumber \\*
 & & \quad + C(\varphi)T^{\rm{DM}}_{ab} + T^{\rm{S}}_{ab},
\end{eqnarray}
where $T^{\rm{DM}}_{ab}$ and $T^{\rm{S}}_{ab}$ are the
energy momentum tensors for (uncoupled) dark matter and standard
model fields. The existence of the scalar field and its coupling
change the form of the energy momentum tensor leading to
potential changes in the background cosmology and 
structure formation.

The coupling to a scalar field produces a direct
interaction (fifth force) between dark matter
particles due to the exchange of scalar quanta. This is best
illustrated by the geodesic equation for dark matter particles
\begin{eqnarray}\label{eq:geodesic}
{{{\rm d}^{2}\bf{r}}\over{{\rm d}t^2}} = -\vec{\nabla}\phi -
{{C_\varphi(\varphi)}\over{C(\varphi)}}\vec{\nabla}\varphi,
\end{eqnarray}
where $\bf{r}$ is the position vector, $t$ the (physical) time, $\phi$
the Newtonian potential and $\vec{\nabla}$ is the spatial
derivative. $C_\varphi\equiv {\rm d}C/{\rm d}\varphi$. The second term in the
right hand side is the fifth force and only exists for coupled matter
species (dark matter in our model). The fifth force also changes the
clustering properties of the dark matter. 

To solve the above two equations we need to know both the time
evolution and the spatial distribution of $\varphi$, {\it i.e.}  we
need the solutions to the scalar field equation of motion (EOM)
\begin{eqnarray}
\nabla^{a}\nabla_a\varphi + {{\rm{d}V(\varphi)}\over{\rm{d}\varphi}} +
\rho_{\rm{DM}}{\frac{\rm{d}C(\varphi)}{\rm{d}\varphi}} = 0,
\end{eqnarray}
or equivalently
\begin{eqnarray}
\nabla^{a}\nabla_a\varphi + {{\rm{d}V_{eff}(\varphi)}\over{\rm{d}\varphi}} =
0,
\end{eqnarray}
where we have defined
\begin{eqnarray}
V_{eff}(\varphi) = V(\varphi) + \rho_{\rm{DM}}C(\varphi).
\end{eqnarray}
The background evolution of $\varphi$ can be solved easily given
the present day value of  $\rho_{\rm{DM}}$ since 
$\rho_{\rm{DM}}\propto a^{-3}$. We can then divide $\varphi$
into two parts, $\varphi=\bar{\varphi}+\delta\varphi$, where
$\bar{\varphi}$ is the background value and $\delta\varphi$ is its
(not necessarily small nor linear) perturbation, and subtract the
background part of the scalar field equation of motion from the full equation
to obtain the equation of motion for $\delta\varphi$. In the
quasi-static limit in which we can neglect time derivatives of
$\delta\varphi$ as compared with its spatial derivatives (which
turns out to be a good approximation on \tcr{galactic and cluster} scales),
we find
\begin{eqnarray}\label{eq:scalar_eom}
\vec{\nabla}^{2}\varphi =
{{\rm{d}C(\varphi)}\over{\rm{d}\varphi}}\rho_{\rm{DM}} -
{{\rm{d}C(\bar{\varphi})}\over{\rm{d}\bar{\varphi}}}\bar{\rho}_{\rm{DM}} +
{{\rm{d}V(\varphi)}\over{\rm{d}\varphi}} -
{{\rm{d}V(\bar{\varphi})}\over{\rm{d}\bar{\varphi}}},
\end{eqnarray}
where $\bar{\rho}_{\rm{DM}}$ is the background dark matter
density.

The computation of the scalar field $\varphi$ using the above equation then completes the computation of the source term for the Poisson equation
\begin{eqnarray}\label{eq:poisson}
\vec{\nabla}^{2}\phi &=& \frac{\kappa}{2}\left[\rho_{\rm tot}+3p_{\rm tot}\right]\nonumber\\
&=& {{\kappa}\over{2}}\left[C(\varphi)\rho_{\rm{DM}} + \rho_{\rm{B}}
- 2V(\varphi)\right], 
\end{eqnarray}
where $\rho_{\rm{B}}$ is the baryon density (we have neglected the kinetic energy of the scalar field because it is always very small for the model studied here).

\subsection{Specification of Model}

\label{subsect:specification}

As mentioned above, to fully fix a model we need to specify the functional forms of $V(\varphi)$ and $C(\varphi)$. Here we will use the models investigated by \cite{lz2009, lz2010, li2011}, with
\begin{eqnarray}\label{eq:coupling}
C(\varphi) = \exp(\gamma\sqrt{\kappa}\varphi),
\end{eqnarray}
and 
\begin{eqnarray}\label{eq:pot_chameleon}
V(\varphi) = {{\Lambda}\over{\left[1-\exp\left(-\sqrt{\kappa}\varphi\right)\right]^\alpha}}.
\end{eqnarray}
In the above $\Lambda$ is a parameter of mass dimension four and 
is of order the present dark energy density 
($\varphi$ plays the role of dark energy in the models). $\gamma, \alpha$ are dimensionless  parameters controlling the strength of the 
coupling and the steepness of the potentials respectively. 

We shall choose $\alpha\ll1$ and $\gamma>0$ as in \citet{lz2009, lz2010},
ensuring that $V_{eff}$ has a global minimum close to $\varphi=0$
and ${\rm d}^2V_{eff}(\varphi)/{\rm d}\varphi^2 \equiv m^2_{\varphi}$ at this
minimum is very large in high density regions. There are two
consequences of these choices of model parameters: (1) $\varphi$ is
trapped close to zero throughout  cosmic history so that
$V(\varphi)\sim\Lambda$ behaves as a cosmological constant; 
(2) the fifth force is strongly
suppressed in high density regions where $\varphi$ acquires a large
mass, $m^2_{\varphi}\gg H^2$ ($H$ is the Hubble expansion rate),
and thus the fifth force cannot propagate far. The suppression of the
fifth force is even stronger at early times, and thus its influence on
structure formation occurs mainly at late times.
The environment-dependent behaviour of the scalar field was
first investigated by \citet{kw,ms}, and is often referred
to as the `chameleon effect'.

\section{First Crossing Probability with Correlated Steps in Chameleon Models}

\label{sect:correlated}

\subsection{Terminology}

In the excursion set approach the calculation of the halo mass function 
${\rm d}n/{\rm d}m$
is mapped to the computation of the first crossing distribution $f(S)$
across some prescribed barriers where
\begin{equation}
\frac{m}{\bar{\rho}} \frac{{\rm d}n}{{\rm d}m} {\rm d}m = f(S){\rm d}S.
\end{equation}
In the following we use the two terms interchangeably. 
The variance of the matter fluctuation field smoothed on scale $R$ is 
given by
\begin{equation}
S = \int \frac{{\rm d}k}{k} \frac{k^3 P(k)}{2\pi^2}W^2(kR),
\end{equation}
where $W$ is the smoothing window function and $P(k)$ is the 
 matter power spectrum linear extrapolated  to the present time.
In hierarchical models $S$ is a monotonic decreasing function of $R$ and 
the smoothing scale in Lagrangian space $R$, 
the total mass enclosed within this scale $M$, and the variance $S$ are
equivalent quantities and can be used interchangeably.

\subsection{Background}

\begin{figure}
\centering\includegraphics[width=0.45\textwidth]{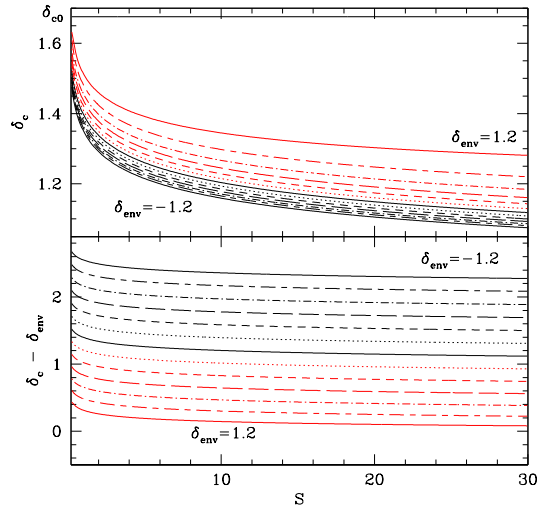}
\caption{Upper panel: Halo formation barrier for various
$\delta_{\rm env}$ from -1.2 to 1.2 with increment 0.2. 
Constant barrier at the top is the case with no fifth force (i.e., $\Lambda$CDM).
Lower panel: The height need to climb to reach the halo formation barrier
for various $\delta_{\rm env}$.}
\label{fig:delc}
\end{figure}

Recent work \citep{le2012,ll2012} demonstrated
how the chameleon-type fifth force modifies the first crossing probability 
and the associated halo mass function for Lagrangian and Eulerian 
environment. This modification is due to the dependence
of the halo formation barrier height $\delta_c(S)$ on the matter density of its surrounding 
environment, $\delta_{\rm env}$\footnote{Note that in this paper we
  use $\delta_{\rm env}$ to denote the initial density perturbation in
  the environment linearly extrapolated to today assuming a
  $\Lambda$CDM model. Knowing $\delta_{\rm env}$ it is straightforward
  to calculate the true environment density in subsequent times using
  the assumed evolution model.}. 
Since the fifth force is strongest when 
the environment density is low, the barrier for halo formation is
lower where $\delta_{\rm env}$ is small.
The upper panel in figure~\ref{fig:delc} shows 
 the halo formation barrier as a function of 
$S$ for various values of $\delta_{\rm env}$
assuming that the spherical collapse model describes halo formation, and confirms that
the fifth force is stronger in underdense environment where
the associated barrier is lower.
The actual height that the random walks with various $\delta_{\rm env}$
need to climb to reach the modified barrier is 
shown in the lower panel -- although the barrier in 
underdense environment is lower, the barrier to overcome is actually higher.

 \citet{le2012,ll2012}, when applying  the excursion set approach 
in chameleon models,
focused on the case where the excursion set is performed with 
uncorrelated steps with two different definitions of environment. 
In both cases the modifications in the first crossing distribution 
in chameleon models are solely due to the change in the barrier.
The situation is more complicated 
when the steps in the random walk are correlated: two additional 
effects can modify the first crossing distribution.
Firstly, the distribution 
of $\delta_{\rm env}$ for Eulerian environments is modified and, 
since the halo formation barrier depends 
on $\delta_{\rm env}$, the first crossing probability will be modified as well.
Note however that when the environment is defined as Lagrangian, 
the distribution of $\delta_{\rm env}$ is unchanged.
Secondly, there will be non-trivial correlations between the 
environment density contrast $\delta_{\rm env}$ and 
the density contrast $\delta$ at different $S$. 

To investigate how these two effects modify the first crossing distribution
in models with chameleon mechanism, we use monte-carlo simulations to 
study the first crossing distribution for three choices of smoothing 
window functions following the procedures in \citet{bcek}.
Our convention is such that 
the normalisation of the power spectrum is set by demanding 
$\sigma_8 = 0.81$ for the tophat filter. 
The sharp-\textit{k} smoothing window function generates random walks with 
uncorrelated steps and it is common practice to use the $S-M$ mapping of the 
tophat window function to relate 
the variance and the smoothing scale for the sharp-\textit{k} smoothing window 
function.
We consider $\Lambda$CDM power spectrum and two environment scales:
an Eulerian environment, which denoted by $\zeta$, is the
  environment at late time; and a Lagrangian environment, which
  denoted by $\xi$, is the environment in the initial conditions. We
  assume both environments are spherical regions that share common
  centers as the proto-halos. The actual scale of the environment should
  be approximately   the Compton wavelength of the scalar field at
  late time. Numerical simulation results suggest that an Eulerian
  scale of $\zeta=5 \sim 8 h^{-1}{\rm Mpc}$ should be chosen
  \citep{lzk2012}. 
In this work we consider  an Eulerian evironment
$\zeta= 5h^{-1}$Mpc and 
a Lagrangian environment 
$ \xi = 8h^{-1}$Mpc, and set
$(\gamma,\alpha)=(0.5,10^{-6})$ as the chameleon model
parameters.
In the language of excursion set formalism, the Lagrangian
  environment barrier (fixing $S$) corresponds to a vertical barrier
  in the $\delta_l-S$ plane while the Eulerian environment barrier is given by the 
spherical collapse model \citep{b1994,sheth98}:
\begin{equation}
b_{\rm Eul}(S) = \delta_{c0} \left[1 -
  \left(\frac{M(S)}{\bar{\rho}\zeta}\right)^{-1/\delta_{c0}}\right],
\end{equation}
where $\bar{\rho}$ is the background density.

\begin{figure}
\centering\includegraphics[width=0.45\textwidth]{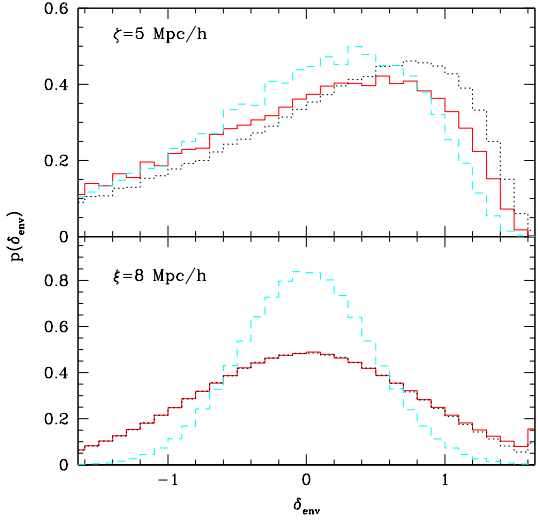}
\caption{(Colour Online) Distribution of environmental density 
 for Eulerian (top panel) and Lagrangian environments (lower panel). The solid, dashed and dotted curves are 
 results of tophat, Gaussian and sharp-{\it k} window functions respectively.
}
\label{fig:pdelenv}
\end{figure}

Figure~\ref{fig:pdelenv} shows the 
distribution of $\delta_{\rm env}$ for $\zeta=5\ {\rm Mpc}/h$ (upper panel)
and $\xi=8\ {\rm Mpc}/h$ (lower panel). 
Three histograms corresponding to tophat (solid), Gaussian (dashed), 
and sharp-\textit{k} (dotted) window functions are shown in each
panel.
We ran a sample of random walk whose correlation satisfies the
  corresponding smoothing filter function. We then record the height
  of the random walk at which it first crosses the environmental
  barrier.  In the case of Lagrangian environment we set $\delta_{\rm
    env}=\delta_{c0}$ if the random walk first crosses $\delta_{c0}$
  on scale bigger than $\xi$.
 Since we use the same $S-M$ mapping for the tophat window
  function and the sharp-$k$ window function, choosing a Lagrangian
  scale fixes $S$ and hence the two will have the same distribution.
The difference  between the distributions of Gaussian smoothing and the others 
in the case of Lagrangian environment is due to our choice of power spectrum 
normalization -- for $\xi=8\ {\rm Mpc}/h$, $\lg(\nu) = 1.1$ for the 
Gaussian filter
while $\lg (\nu)=0.62$ for the tophat and sharp-\textit{k} filters. Here $\nu\equiv\delta_{\rm c0}^2/S$.
Since the Lagrangian $\delta_{\rm env}$ distribution is the same for 
the tophat and sharp-\textit{k} filters, any difference in their 
first crossing distributions due to the fifth force has to 
come from the non-trivial correlations between the height of the random walk
and $\delta_{\rm env}$ (see below).
On the other hand, the distributions in Eulerian environment are different 
for all three smoothing windows: the sharp-\textit{k} filter peaks at highest 
$\delta_{\rm env}$ while the Gaussian filter peaks at lowest $\delta_{\rm env}$.
As a result the effect of the fifth force is stronger when the Gaussian 
window function is used -- 
the halo formation barrier for walks with Gaussian filter is lower than that of 
uncorrelated walks. 
Analytical approximations to evaluate the first crossing of 
Eulerian environment are available for both uncorrelated 
\citep{zh2006,lamsheth09}
and correlated steps \citep{ms12}.

\subsection{Monte-Carlo simulations}

\begin{figure}
\centering\includegraphics[width=0.5\textwidth]{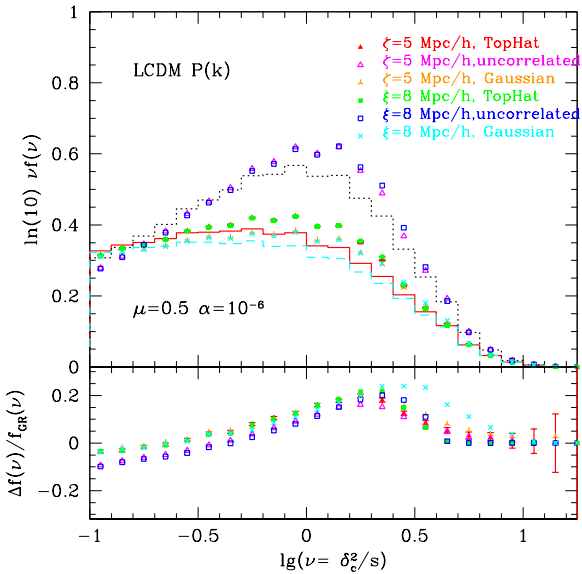}
\caption{(Colour Online) Comparison of the first crossing distribution 
in chameleon models for correlated as well as uncorrelated steps. 
Three sets of histograms are the first crossing probability 
for the $\Lambda$CDM model, a constant 
barrier $\delta_c=1.676$, for different smoothing window functions:
from top to bottom are sharp-\textit{k} filter (uncorrelated walks), 
tophat filter, and Gaussian filter.
The symbols are the first crossing distribution when there is the fifth force for $\zeta = 5h^{-1}$Mpc and $\xi = 8h^{-1}$Mpc.
See the legends for the details.
The bottom panel shows the fractional change in the first crossing probability
with respect to the corresponding constant barrier first crossing probability.
Only error bars for red solid triangles are included for 
clarity. }
\label{fig:nufnuCorr}
\end{figure}

The top panel in figure~\ref{fig:nufnuCorr} shows the first crossing distributions
in chameleon models for both correlated and uncorrelated walks.
The histograms show the distribution when there is no fifth force: 
it is the first crossing for a constant barrier at $\delta_c= 1.676$.
The line labels are the same as in figure~\ref{fig:pdelenv}.
Symbols represent the first crossing distributions 
for different combinations of smoothing window functions and choices 
of environment, as indicated in the legends. 
The first crossing distributions change significantly 
when the random walks switch from uncorrelated steps (sharp-\textit{k}) 
to the correlated one (Gaussian and tophat); 
different definitions of the environment have subdominant effect. 
To better visualise the effect of different choices of the environment,
the lower panel shows the relative difference in the distribution to the 
corresponding constant barrier distribution.

The modifications in the first crossing distribution due to 
the chameleon-type fifth force are quite different for 
random walks with correlated steps (both Gaussian and tophat filters) 
versus walks with uncorrelated steps.
For example the modifications of the correlated steps are displaced 
from those of the uncorrelated steps for $\lg(\nu) < 0.1$.
Since the modification in the halo mass function is around 20\%,
this $\sim5$\% displacement is a big contribution.
Consider for example the tophat 
(solid squares) and the sharp-\textit{k} (open squares) 
Lagrangian environments --
since they have the 
same distribution $p(\delta_{\rm env})$, the difference in the predicted 
modification in the first crossing distributions must be due to the 
nontrivial correlations between different smoothing scales when the window 
function is tophat.
One can imagine the following scenario:
random walks with correlated steps are less likely to have dramatic 
fluctuations compared to those with uncorrelated steps -- 
when $\delta_{\rm env} < 0$, the enhanced 
fifth force lowers the barrier $\delta_c$ --
in both uncorrelated and correlated cases, it is easier to cross 
the barrier. However correlated walks are not likely have sudden 
jumps, hence the modification in the first crossing comes only at bigger $S$:
it explains the difference between the solid (tophat Lagrangian) and open 
(sharp-\textit{k} Lagrangian) symbols for $\lg(\nu) > 0.3$ 
(Lagrangian scale $\xi=8\ {\rm Mpc}/h$ corresponds to $\lg(\nu)=0.62$).
For bigger $S$ the correlated walks do not need have sudden 
jumps to cross the barrier --
hence the increase in the first crossing probability. 
On the other hand, the increase for the uncorrelated walks is 
smaller since some of those walks 
already cross the barrier at smaller $S$.
For $\delta_{\rm env} > 0$, the fifth force is weak and only slightly 
enhances structure formation. 
In this case walks with correlated steps may upcross the barrier 
sooner than walks with uncorrelated steps 
(since uncorrelated walks can have sudden decrease in height). 
It will result in higher probability in first crossing distribution 
for correlated steps for $S$ near the chameleon environment -- 
however this modification in the first crossing distribution relative to the 
GR case is small due to the weakening of the fifth force.
Recall that it is equally likely to have overdense and
  underdense environment in Lagrangian environment, 
the net effect is correlated steps show a smaller change in first
crossing probability for small $s$ but bigger change for big $s$.


The change in first crossing distribution for Gaussian filter and
Lagrangian environment is strong in high $\nu$. It is consistent with 
the change in the distribution of $\delta_{\rm env}$ shown in 
figure~\ref{fig:pdelenv}: since $p(\delta_{\rm env})$ has a narrower 
distribution with a peak at $\delta_{\rm env}=0$, 
the overdense environments in the Gaussian case generally have lower $\delta_{\rm env}$ and so the fifth force is 
stronger (lower barrier) compared to the walks with tophat or 
sharp-\textit{k} filter. 
On the other hand the change at low mass end is similar to 
others -- although there are more walks that have underdense 
Lagrangian environments for the other two filters,
 those walks need to climb a huge distance to cross their respective 
modified barriers and do not make a significant contribution in the first
crossing distribution.

\subsection{Analytical approximation}\label{sect:analytical}

In this subsection we apply the analytical approximation proposed
by \citet{ms12} to estimate the first crossing distribution in chameleon models.
In this analytical approximation the first crossing probability at $s$
is approximated by both the height and its rate of change at $s-\Delta s$.
The first crossing distribution across barrier $b(s)$
\footnote{In this subsection we use upper and lower letters to denote 
different smoothing scales and random walk heights.} 
is 
\begin{equation}
f(s) = p(b,s)\int_{b'}^{\infty}{\rm d}\delta' \ p(\delta'|b,s) (\delta' - b'),
\end{equation}
where the dash denotes derivative with respect to $s$, 
$p(\delta'|b,s)$ is the conditional probability density of the rate of 
change in height $\delta'$ at $s$ given
that  the walk's height is $b$ at the same scale $s$, 
and $p(\delta,s)$ is the probability density that the walk has height $b$ 
at the smoothing scale $s$.

When the Lagrangian environment is used in the chameleon models, 
the first crossing distribution of the modified barrier $b=\delta_c(s,\delta_{\rm env})$ at $s$
is 
\begin{eqnarray}
f_{\rm Lag}(s;\xi) &= & \int_{-\infty}^{\delta_{c0}} {\rm d}\delta_{\rm env} \ 
    p(\delta_{\rm env},\xi) p(b,s|\delta_{\rm env},\xi) \nonumber \\
   & &\times   \int _{b'}^{\infty}{\rm d}\delta' \ p(\delta'|b,s;\delta_{\rm env},\xi) (\delta' - b'),
\label{eqn:fnuLag}
\end{eqnarray}
where $p(\delta_{\rm env},\xi)$ is the distribution 
of environmental density contrast at scale $\xi$. 
Strictly speaking $p(\delta_{\rm env},\xi)$ should be replaced by 
the condition probability that $\delta$ never crosses $\delta_{c0}$ 
for scales bigger than $\xi$, but the scales we apply the Lagrangian 
environment is big enough so that the difference between the two is small.

\begin{figure}
\centering\includegraphics[width=0.5\textwidth]{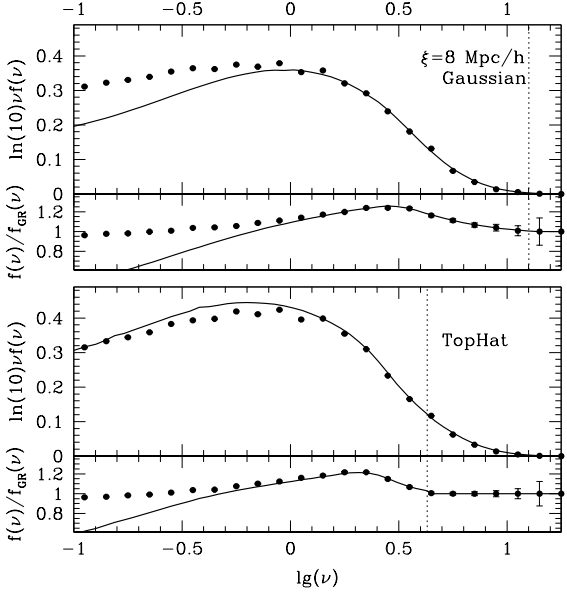}
\caption{Analytical approximation for first crossing distribution in chameleon 
models for correlated steps. The upper and the lower figure show the 
results using Lagrangian environment $\xi=8\ {\rm Mpc}/h$ 
for Gaussian and tophat window filters respectively. 
In each figure the curve is the analytical approximation while the data points 
are monte-carlo simulation results. The upper half of each figure shows 
the first crossing distribution and the lower half shows the ratio to the 
constant barrier distribution. The vertical dotted lines show the scales 
of the Lagrangian environment.}
\label{fig:nufnuLag}
\end{figure}

Figure~\ref{fig:nufnuLag} shows the comparison of the analytical 
approximation and the monte-carlo results for Lagrangian environment of
$\xi=8{\rm Mpc}/h$. The 
upper panel shows the results of Gaussian window function while the 
lower panel shows the tophat window function.
In each panel the top half shows the first crossing distribution from 
analytical prediction (curve) and the monte-carlo result (symbols), and
the bottom half shows the modifications to the first crossing distribution 
relative to GR.
The analytical approximation matches the monte-carlo simulation for 
$\lg(\nu) > 0$ -- for smaller $\nu$, multiple crossings are more frequent
and the approximation breaks down.

\begin{figure}
\centering\includegraphics[width=0.5\textwidth]{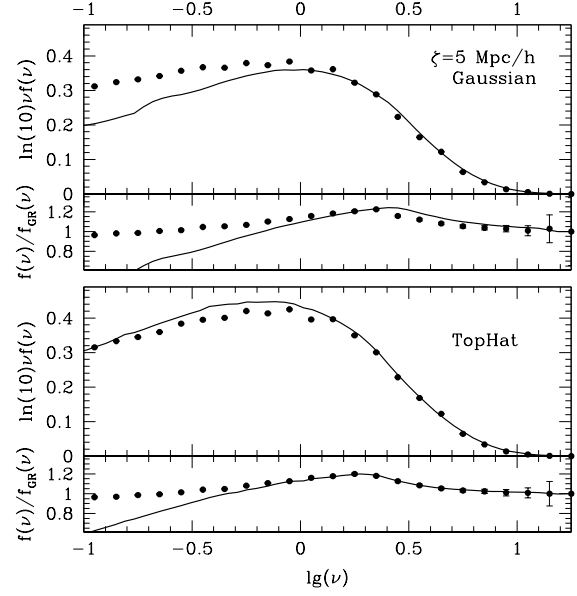}
\caption{Similar to figure~\ref{fig:nufnuLag} but using Eulerian 
environment $\zeta = 5\ {\rm Mpc}/h$.}
\label{fig:nufnuEul}
\end{figure}

The first crossing distribution for Eulerian environment contains two first
crossings: first to cross the environment barrier, 
then the halo formation barrier. Applying the approximation to both crossings 
yields
\begin{eqnarray}
f_{\rm Eul}(s,\zeta)& = & \int_0^s {\rm d}S\ p(B,S)
          \int_{B'}^{\infty}{\rm d}\Delta'\ p(\Delta'|B,S)(\Delta'-B')\nonumber \\*
  &  &\times       p(b,s|B,S;\Delta') \nonumber \\*
   & & \quad \times \int_{b'}^{\infty}{\rm d}\delta'\ 
               p(\delta'|b,s;B,S;\Delta') (\delta' - b'),
\label{eqn:fnuEul}
\end{eqnarray}
where we denote the Eulerian environment barrier as $B(S)$ and 
$\Delta'$ is the derivative of $\delta_{\rm env}$ at $S$. 
The dependence of the halo barrier $b(s)$ on $B(S)$ is implicit.
The first integral includes walks with different Eulerian 
environment density contrast (up to $s$)
while the second integral corresponds to the 
first crossing of the Eulerian barrier at $S$.
Here one should have replaced the upper limit of the integration
such that
the walk would not cross $\delta_{c0}$ at $S$ -- in practice the probability 
of having a big jump is not likely so using the above approximation 
does not alter the result. 
Finally the rest corresponds to the first crossing of the modified halo barrier 
at $s$ given the walk first crossed the Eulerian barrier at $S$ with 
an increase in height $\Delta'$ when going from $S-\Delta S$ to $S$.
The results are shown in figure~\ref{fig:nufnuEul}.
The analytical approximation works relatively well for 
big $\nu$ 
($>1$) but breaks down for small $\nu$, 
as in the case of Lagrangian environment.

The conditional probability distributions in this subsection are also Gaussian
distributed with mean and variance given by
\begin{eqnarray}
{\rm E}(x|y_1,y_2,\cdots,y_n)  &= & 
           \sum_{i,j=1}^n \langle xy_i\rangle A_{ij}^{-1}y_j  \\
{\rm Var}(x|y_1,y_2,\cdots,y_n) & = & {\rm Var}(x) - \sum_{j=1}^n \langle xy_j \rangle^2 A_{jj}^{-1}\nonumber \\*
  & & - 2 \sum_{j>k} \langle xy_j \rangle \langle xy_k\rangle A_{jk}^{-1},
\end{eqnarray}
where $A$ is the covariance matrix of $\{y_i\}$ and we assume all 
variables have unconditional means vanish in the above formuale.

\section{Conditional Mass Function and Halo Bias in Chameleon Models}

\label{sect:condition}

\begin{figure*}
\centering
  \subfloat[$\delta_0 =\pm 0.35$]{
      \label{fig:connmass0.35b0.15}
      \includegraphics[width=0.44\linewidth]{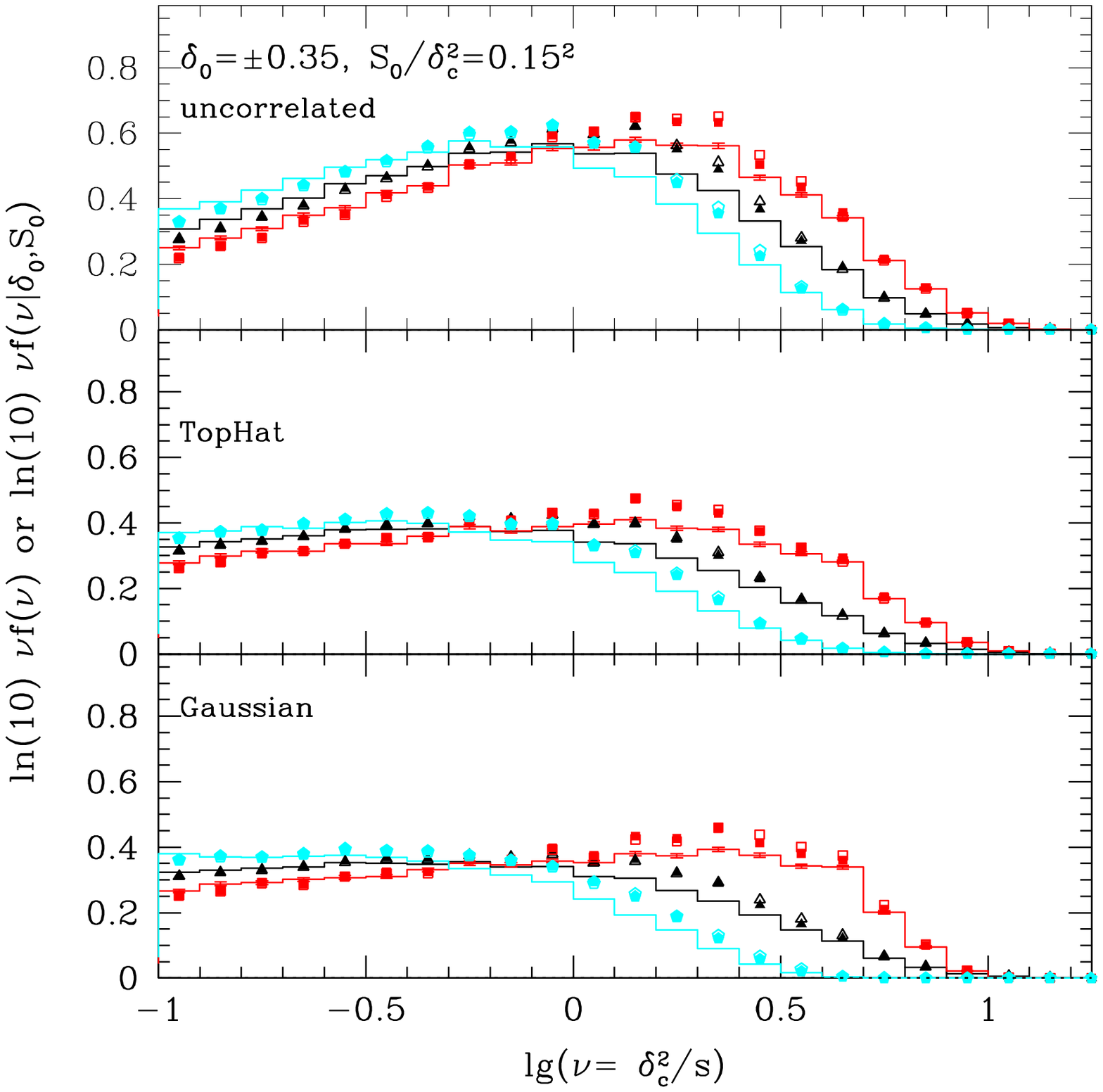}}
   \hspace{0.07\linewidth}
  \subfloat[$\delta_0 =\pm 0.7$]{
      \label{fig:connmass0.7b0.15}
      \includegraphics[width=0.44\linewidth]{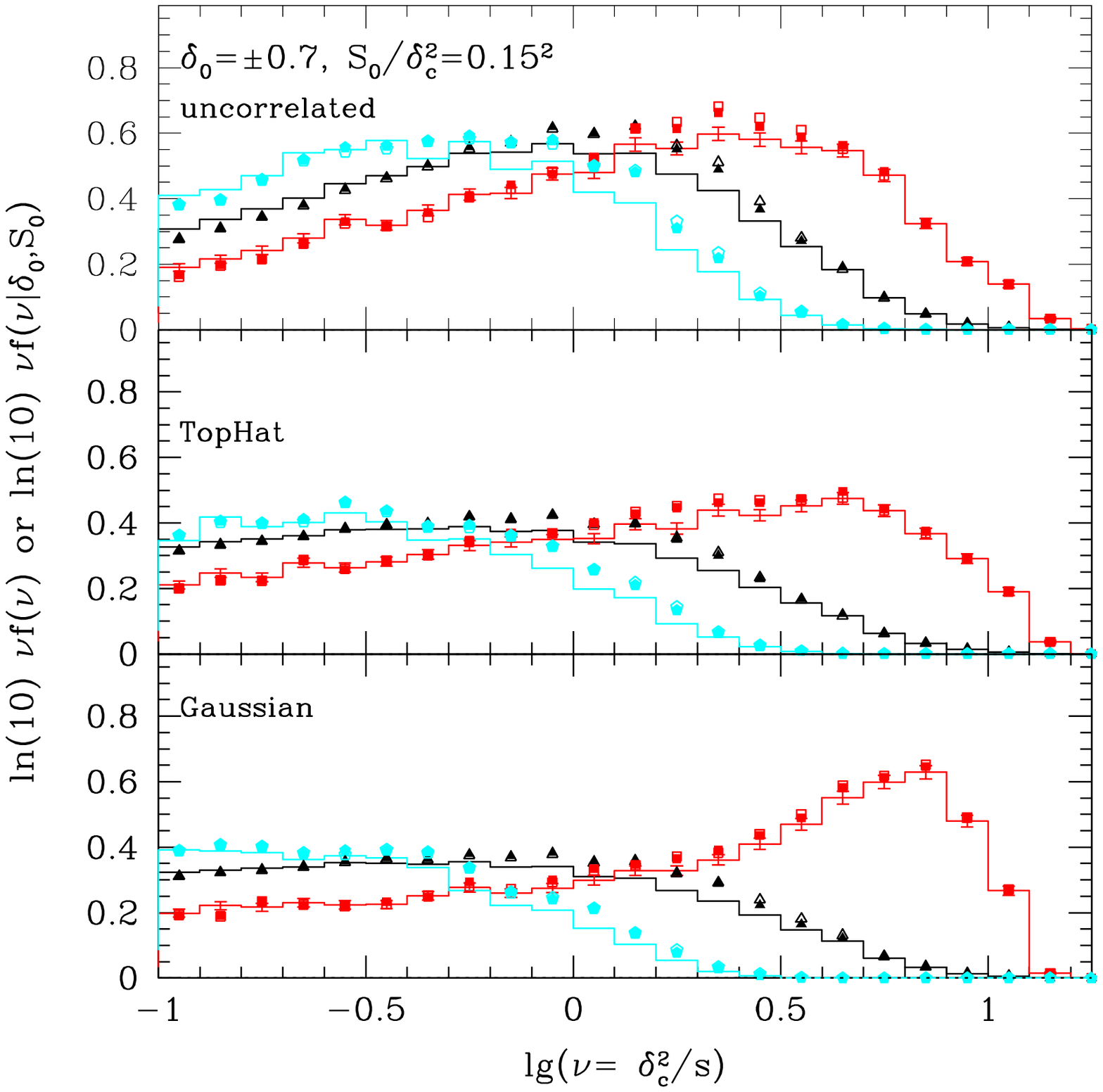}}
\caption{(Colour Online) 
Comparison of the conditional 
first crossing probability
obtained from Monte Carlo simulations for uncorrelated and correlated steps
by selecting subset of walks that have various values of $\delta_0$ at
$S_0 = 0.15^2\delta_c^2$.
In each panel the histograms indicate the first crossing
probability for the constant barrier case ($\Lambda$CDM) while 
the the solid symbols and the open symbols represent 
chameleon models with $\zeta = 5h^{-1}$Mpc and $\xi = 8h^{-1}$Mpc
respectively. 
Three sets of probability are included in each panel: $|\delta_0|$ 
(highest amplitude histogram at $\lg(\nu)=0.5$ and squares); 
$-|\delta_0|$ (lowest amplitude histogram at $\lg(\nu)=0.5$  and pentagons); 
and unconditional probability (the intermediate histogram and triangles).
Different panels use different smoothing window function:
(from top to bottom): uncorrelated steps; tophat window function; Gaussian 
window function. Only error bars for the histogram for positive $\delta_0$
are included for clarity.
}
\label{fig:connmassb0.15}
\end{figure*}

\begin{figure*}
\centering
  \subfloat[$\delta_0 =\pm 0.5$]{
      \label{fig:connmass0.5b0.25}
      \includegraphics[width=0.44\linewidth]{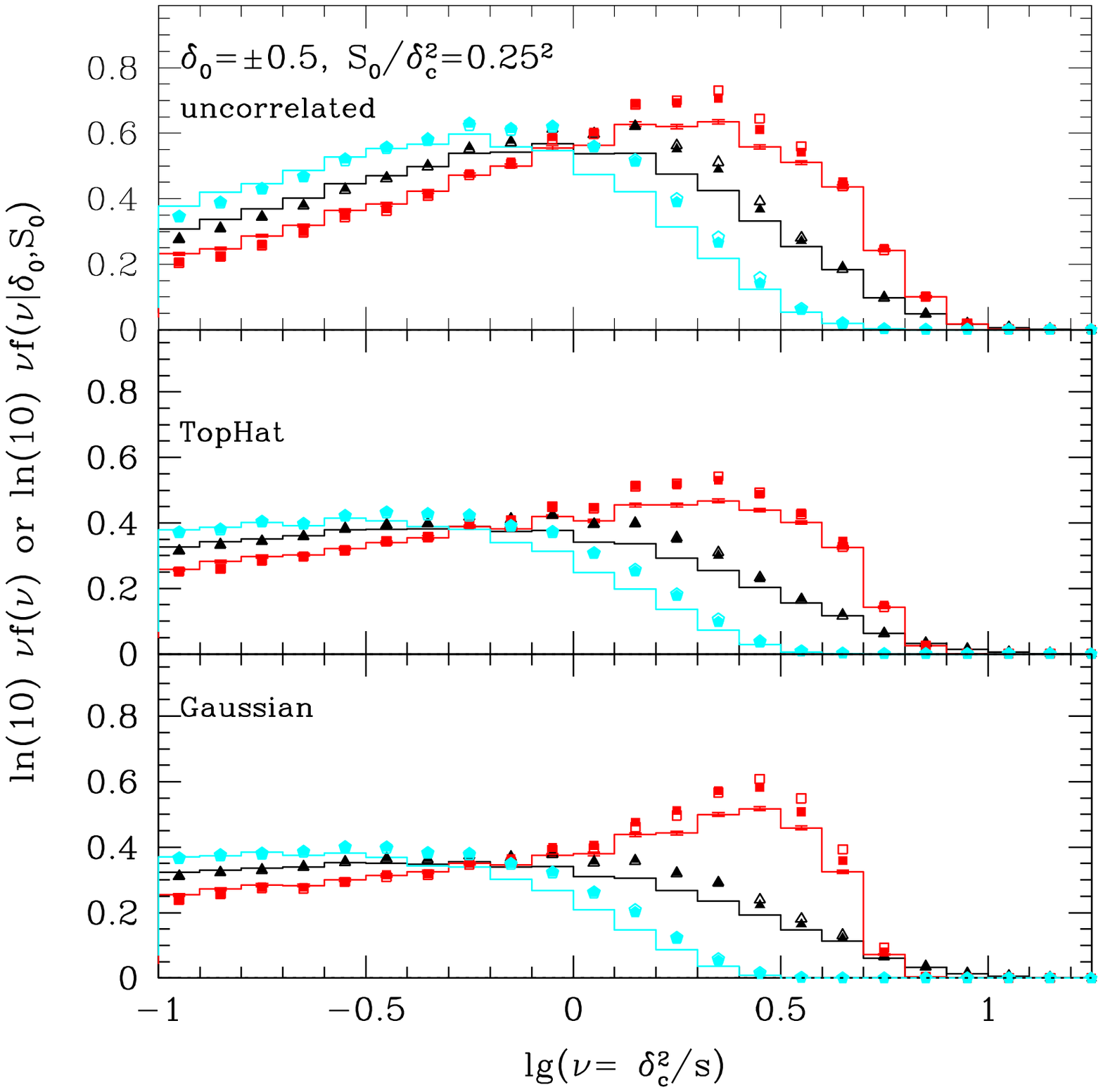}}
   \hspace{0.07\linewidth}
  \subfloat[$\delta_0 =\pm 0.9$]{
      \label{fig:connmass0.9b0.25}
      \includegraphics[width=0.44\linewidth]{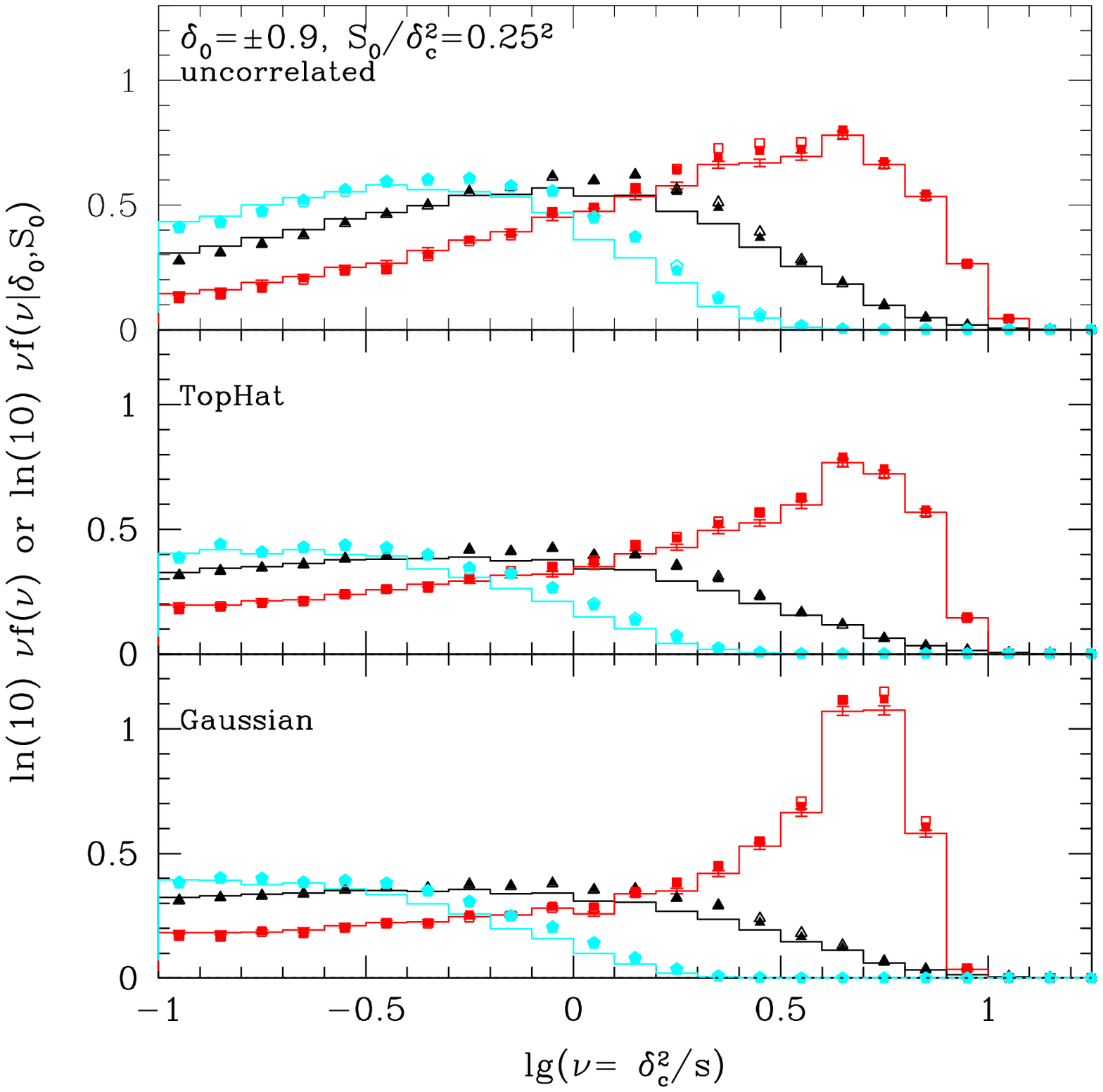}}
\caption{(Colour Online) Conditional first crossing probability similar to 
 figure~\ref{fig:connmassb0.15} the condition is set at $S = 0.25^2\delta_c^2$.}
\label{fig:connmassb0.25}
\end{figure*}

In this section we will describe the effect of the chameleon-type fifth force on 
the conditional mass function and the halo bias. 
As explained in the previous section, the fifth force 
is stronger when the environment is less dense -- effectively 
the halo formation criterion is lower and hence 
the first crossing of halo formation barrier is easier.
One may then ask if this effect will be more significant when 
we look at the conditional mass function, especially for underdense 
regions. Here we will use Monte Carlo simulations to investigate
how the conditional mass function responses to the fifth force
introduced by the chameleon models.

We now have two environments: 
one is used to solve the fifth force in the chameleon model 
(we will call it the {\it chameleon} environment hereafter and we choose $\zeta= 5h^{-1}$Mpc and 
$\xi = 8h^{-1}$Mpc) and the other is a {\it large-scale} environment 
 $S=S_0$ on which the condition is defined. We will choose $S_0$ to be small (hence very large
scale) so that it is safe to assume that the range of the fifth force
is much smaller than the scale corresponding to $S_0$.

\subsection{Conditional first crossing probability in chameleon models}

In this subsection we will look at the conditional first crossing probability
for three different smoothing window functions.
The conditional first crossing probability is 
\begin{equation}
f(S|\delta_0,S_0) = \int {\rm d}\delta_{\rm env}\ f(S|\delta_{\rm env};\delta_0,S_0)
 P(\delta_{\rm env}|\delta_0,S_0),
\end{equation}
where $f(S|\delta_{\rm env};\delta_0,S_0)$ is the conditional 
first crossing probability 
across the $\delta_c(S,\delta_{\rm env})$ barrier given that it 
first crosses the environment barrier $B(s)$ at $\delta_{\rm env}$ \textit{and}
it has $\delta=\delta_0$ at $S_0$. $P(\delta_{\rm env}|\delta_0,S_0)$ is 
the conditional first crossing probability of crossing the environment barrier
at $\delta_{\rm env}$. 
Changing the large scale environment can have two effects: 
\begin{enumerate}
\item It modifies the distribution of $\delta_{\rm env}$: 
      when $\delta_0 < 0$, it is more likely to have less dense 
      chameleon environment ($\delta_{\rm env}$ is smaller) and 
      hence the fifth force will be stronger.
\item It also affects the first crossing distribution of 
      the halo formation barrier when the random walk is non-Markovian 
      (this effect is additional to the first point where the halo formation
       barrier is changed due to the change of environment distribution).
\end{enumerate}
One may expect the second effect to be weaker than the first since 
the correlation is weaker for bigger difference in $S$. 
In particular, for sharp-\textit{k} filter where 
the random walks have uncorrelated steps, 
the walks are Markovian and 
the large scale environment $(\delta_0,S_0)$ only modifies 
the distribution of $\delta_{\rm env}$, that is,
\begin{equation}
f_{\rm uncor}(S|\delta_{\rm env};\delta_0,S_0) = f(S|\delta_{\rm env}).
\end{equation}
We are going to quantify the effect in this section using monte-carlo 
simulations. We choose two large scales 
($S_0= 0.15^2\delta_{c0}^2$ and $0.25^2\delta_{c0}^2$) and measure
the conditional first crossing distributions using different window filters.

Figures~\ref{fig:connmassb0.15} and~\ref{fig:connmassb0.25} show the
conditional first crossing distributions for various choices 
of $\delta_0$ at $S_0 =0.15^2\delta_c^2$ and  $0.25^2\delta_c^2$ respectively. 
In each panel the histograms indicate the first crossing
probability for the constant barrier case ($\Lambda$CDM), while 
the solid and the open symbols represent 
chameleon models with $\zeta = 5\ {\rm Mpc/h}$ and $\xi = 8\ {\rm Mpc}/h$ respectively. 
Three sets of distribution are included in each panel: 
conditional probability with $\delta_0>0$
(highest amplitude histogram at $\lg(\nu)=0.5$ and squares),
conditional probability with $\delta_0<0$ (lowest amplitude histogram at $\lg(\nu)=0.5$ and pentagons),
and unconditional probability (the intermediate histogram and triangles).
Different panels use different smoothing window functions, from top to bottom:
sharp-\textit{k}, tophat, and Gaussian filters.
The conditional mass function differs from the unconditional mass 
function and the change depends on smoothing window function as well as
the large scale density contrast $\delta_0$: in general 
conditional distributions using Gaussian or tophat window functions and 
having more extreme values of $\delta_0$ show more significant changes compared
to the respective unconditional distribution. 

\begin{figure}
\centering\includegraphics[width=0.5\textwidth]{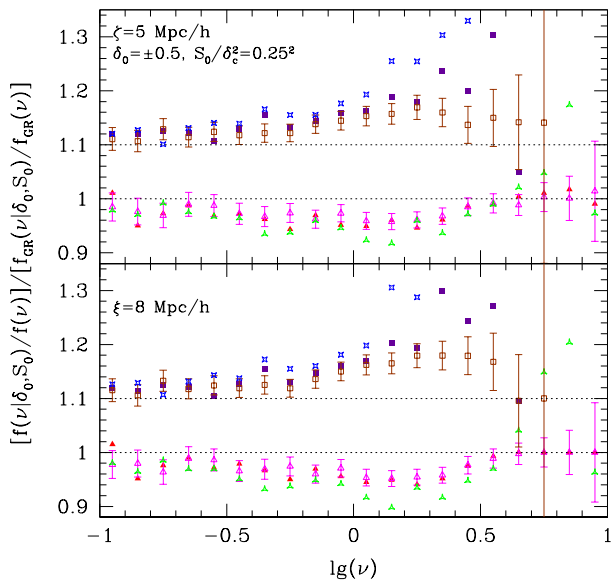}
\caption{(Colour Online)
 Comparison of how the change in first crossing probability due to 
 the fifth force depends on large scale environment $(\delta_0,S_0)$.
 The ratio are taken from the 
 conditional distributions of figure~\ref{fig:connmass0.5b0.25}
  and compared to ratio taken from the unconditional distributions.
 The upper panel shows  results with Eulerian environment 
$\zeta=5\ {\rm Mpc}/h$ and the lower panel is Lagrangian environment
$\xi=8\ {\rm Mpc}/h$. Two sets of points in each panel indicate 
$\delta_0=0.5$ (lower set) and $\delta_0=-0.5$ (upper set, shifted by +0.1)
respectively. Each set has three symbols representing different 
window functions: open (uncorrelated walks), solid (tophat), starred (Gaussian).
Only error bars for uncorrelated walks are included as an approximate 
error indicator.
 }
\label{fig:connmassratio0.5b0.25}
\end{figure}

\begin{figure}
\centering\includegraphics[width=0.5\textwidth]{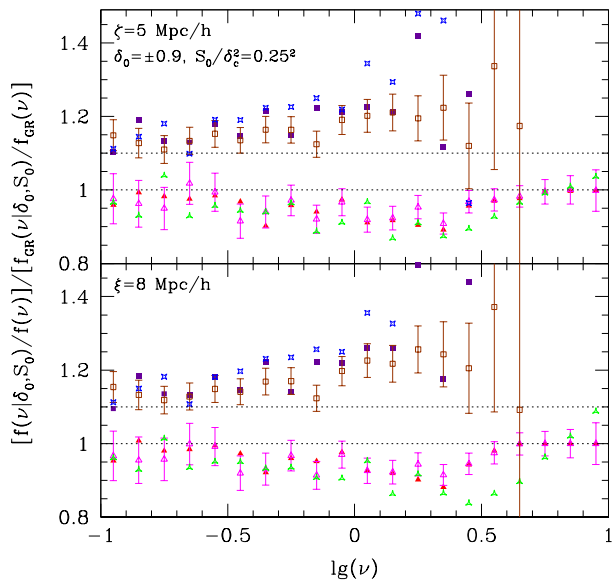}
\caption{(Colour Online) Conditional first crossing probability similar to 
 figure~\ref{fig:connmassratio0.5b0.25} (note the change in y-axis range) 
but the subsets of walks pass through 
$\delta_0=\pm 0.9$.}
\label{fig:connmassratio0.9b0.25}
\end{figure}

The large scale  environment modifies the distribution of the linear density 
contrast at the chameleon environment, which 
 in turn modifies the first crossing distribution.
The modification depends on the smoothing window functions due to the 
difference in correlation strength between various scales involved 
(the large scale environment, 
the chameleon environment, as well as the scale at which the halo barrier 
is first crossed).
We took the ratio of the change in first crossing distribution
for conditional walks 
to that of the unconditional walks and the results are 
shown in figure~\ref{fig:connmassratio0.5b0.25} and 
\ref{fig:connmassratio0.9b0.25}
for $\delta_0 = \pm 0.5$ and $\pm 0.9$ respectively 
when $S_0=0.25^2\delta_{c0}^2$. 
The upper and lower panels of the two figures show respectively 
Eulerian environment and Lagrangian environment for the chameleon 
models. 
In each panel the ratios for $\delta_0 < 0$ are shifted upwards by 0.1 
and  symbols represent different smoothing window kernels:
open symbols for sharp-\textit{k}, solid symbols for tophat, and starred
symbols for Gaussian.

If the change in the first crossing probability 
induced by the chameleon-type fifth force did not depend on 
the large scale environment, 
this ratio should be unity (indicated by the 
dotted horizontal line).
The ratio above (below) unity indicates that
those walks experience stronger (weaker) chameleon effect 
compared to the unconditional walks.
They also make relative
bigger (smaller) contributions to the change in 
the unconditional first crossing distribution.
The monte-carlo results are consistent with expectations:
for walks that have slightly overdense large scale environment ($\delta_0>0$), 
the ratio is always less than unity -- 
these walks are more likely to have overdense $\delta_{\rm env}$ 
(regardless of the choice of  Eulerian or the Lagrangian environment) and 
hence the strength of the fifth force is weaker.
The ratios for walks with correlated steps (solid and starred triangles) 
have bigger deviations from unity compared to 
the ratio for walks with uncorrelated steps (open triangles), 
indicating the conditional distribution of $\delta_{\rm env}$ 
has stronger correlations with $\delta_0$ when steps are correlated 
(otherwise the correlation between $\delta_0$ and $\delta$ at some bigger 
$s$ should result in more first crossing).
When the value of $\delta_0$ is more extreme, 
the deviations from unity is bigger
indicating further weakening of the fifth force.

On the other hand 
for walks that have underdense large scale environment ($\delta_0 < 0$), 
the ratio for both uncorrelated and correlated walks 
stays above unity: the change in 
the first crossing distribution due to the fifth force is
stronger than the unconditional walks. 
The ratio goes significantly above unity for correlated
steps -- indicating the abundance of intermediate mass halos in underdense 
environment would potentially provide a strong constraint on the chameleon-type fifth force.

\begin{figure}
\centering\includegraphics[width=0.5\textwidth]{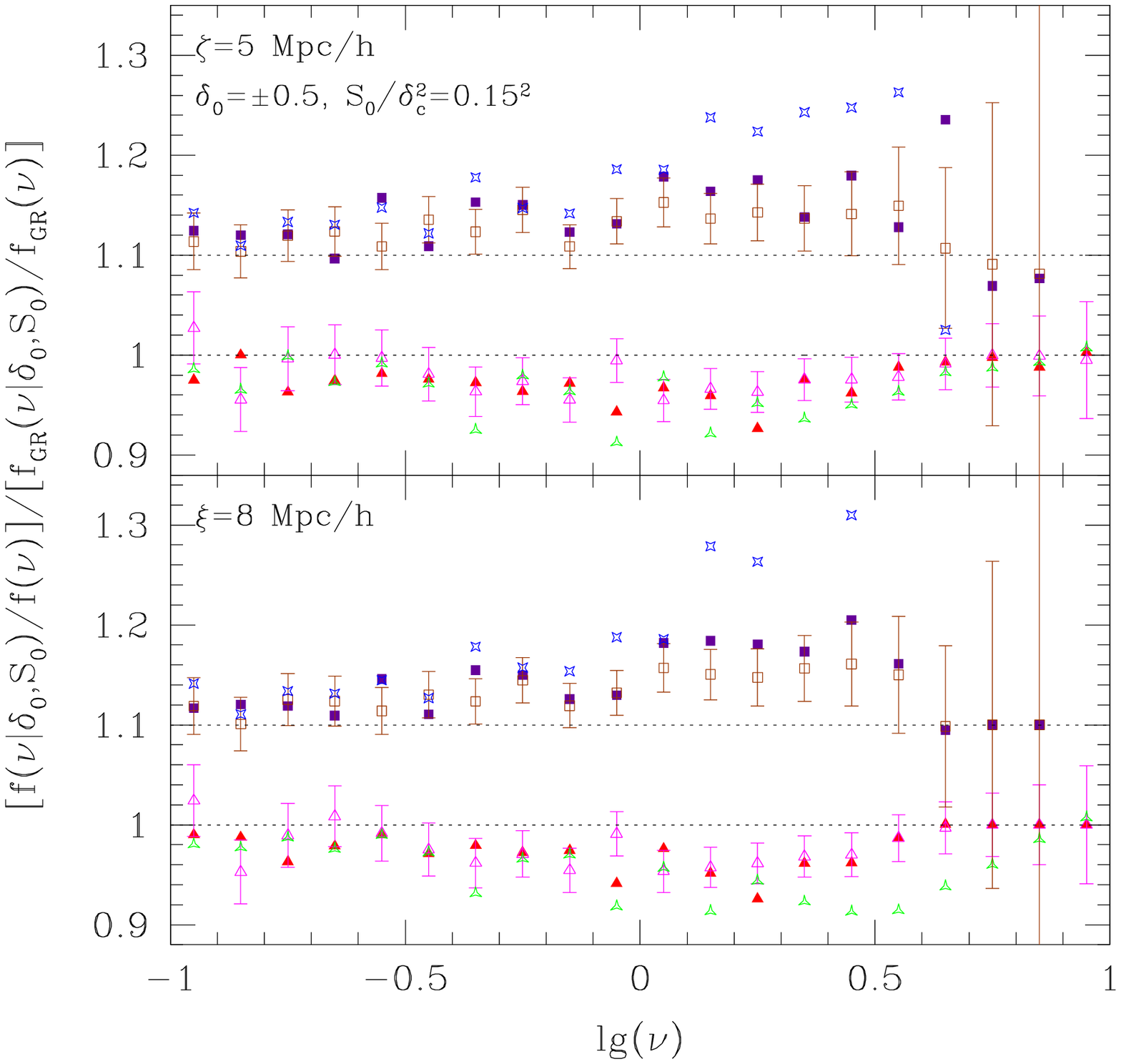}
\caption{(Colour Online) Conditional first crossing probability similar to 
 figure~\ref{fig:connmassratio0.5b0.25} 
but the subsets of walks pass through 
$\delta_0=\pm 0.5$ at $S_0=0.15^2\delta_{c0}^2$.}
\label{fig:connmassratio0.5b0.15}
\end{figure}

Figure~\ref{fig:connmassratio0.5b0.15} shows a similar plot but with the large scale 
environment set at $S_0=0.15^2\delta_{c0}^2$. Comparison to 
figure~\ref{fig:connmassratio0.5b0.25} 
indicates that while the conditional distribution $p(\delta_{\rm env}|\delta_0,S_0)$ 
still differs from the unconditional distribution 
$p(\delta_{\rm env})$, the difference is less significant -- 
very large environment ($S_0=0.15^2\delta_{c0}^2$) 
has weaker correlations with the chameleon 
environment compared to the previous case where  $S_0=0.25^2\delta_{c0}^2$.

In this subsection we have studied the conditional mass function in chameleon models.
The correlation  between the large scale environment and the chameleon 
environment induces further modifications in the first crossing distribution --
this effect is strongest (in terms of the modification of the conditional 
mass function) when the large scale environment is underdense. 
The corresponding change in the halo mass function for intermediate mass halo
shows an extra $\sim$30\% boost compared to the unconditional mass function.
It is straightforward to extend the analytical approximation in 
\S~\ref{sect:analytical} to obtain the conditional first 
crossing distribution by including an additional condition $(\delta_0,S_0)$ 
in all the probability distributions in equations~\eqref{eqn:fnuLag} 
and \eqref{eqn:fnuEul}. This is beyond the scope of this work. 

\subsection{Halo bias in chameleon models}

Halo bias describes the relationship between the halo 
overdensity $\delta_h$ and the matter overdensity $\delta$ where
the halo overdensity can be expanded as  
\begin{equation}
\delta_h = \sum_{i=1} \frac{b_i}{i!}\delta^i.\label{eqn:bias}
\end{equation}
On large scale the above expansion can be truncated in the first few orders.
In particular the linear bias term $b_1$ is commonly used to fit the ratio 
between the halo-matter power spectrum $P_{h\delta}$ 
and the matter power spectrum $P_{\delta\delta}$ 
(or its square as the ratio between halo power spectrum $P_{hh}$
and $P_{\delta\delta}$).
The excursion set approach provides a natural way to relate
halo abundance and halo bias -- by starting the random walks at 
some prescribed locations rather the origin \citep{mw1996,st1999}. 
The relative difference of this conditional first crossing probability to 
the unconditional one  
gives the left hand side of 
equation~\eqref{eqn:bias}. 

Recently there are several studies \citep{maetal11,ps2012} on 
evaluating the halo bias using the excursion set approach with correlated
steps. Both of the analyses (using very different methods) 
found that the halo bias with correlated steps is stronger than
the uncorrelated case. \citet{ps2012} also suggests that, when correlated
steps are considered, the 
halo bias computed from the excursion set approach is different from the 
one obtained by taking the ratio of the halo-matter power spectrum and 
the matter power spectrum. 
While we are interested in computing the halo bias with correlated steps,
we do not make the distinction here and we will present 
the halo bias evaluated within the excursion set framework.

\begin{figure}
\centering\includegraphics[width=0.5\textwidth]{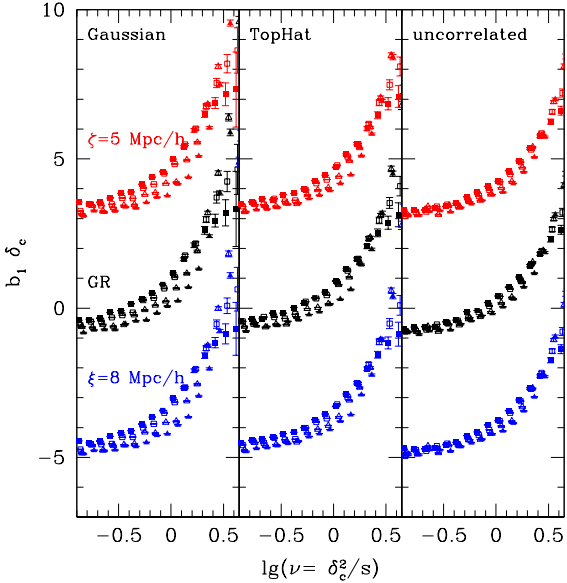}
\caption{(Colour Online) Halo bias computed from monte-carlo simulations
         using equation~\eqref{eqn:b1}. We chose $S_0 =0.15^2\delta_{c0}^2$
         and $\delta_0=\pm 0.35 \ {\rm (solid)\ or}\ \pm 0.5\ {\rm (open)}$ 
         for the conditional first crossing distribution where
         triangles (squares) represent positive (negative) $\delta_0$.
}
\label{fig:b1}
\end{figure}

In the chameleon models the halo formation barrier in the excursion set 
approach depends on the environment density. 
To estimate the halo bias we use the conditional first crossing distributions
obtained in the previous subsection and compute the halo bias from 
equation~\eqref{eqn:bias} where 
\begin{equation}
b_1\delta_{c0} = \left[\frac{f(S|\delta_0,S_0)}{f(S)} -1 \right]\frac{\delta_{c0}}{\delta_0}, \label{eqn:b1}
\end{equation}
assuming high order terms can be neglected.
We chose the conditional first crossing distribution where 
$S_0=0.15^2\delta_{c0}^2$ and $\delta_0 =\pm 0.35$ and $\pm 0.5$ 
and the results are shown in figure~\ref{fig:b1}.
The three panels show results using different smoothing window functions
and the results for chameleon models are shifted by $\pm 4$.

\begin{figure}
\centering\includegraphics[width=0.5\textwidth]{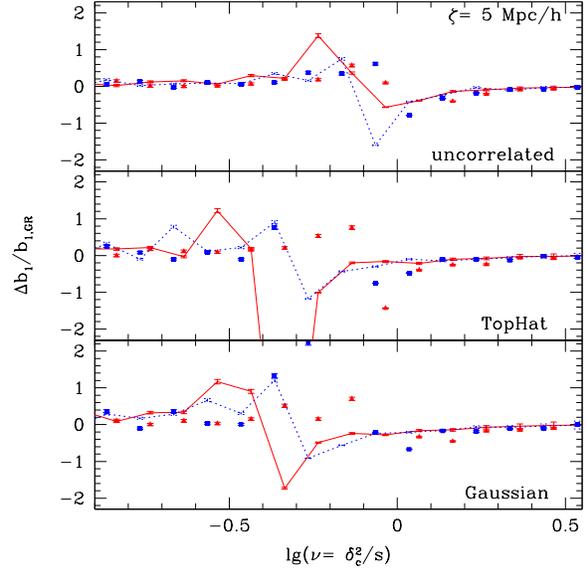}
\caption{(Colour Online) Relative difference in halo bias 
         for chameleon models (Eulerian environment) to that of GR. 
         Triangles and solid lines are the relative difference for 
         $\delta_0=\pm 0.5$ respectively; squares and dotted lines 
        for $\delta_0=\pm 0.35$.
}
\label{fig:b1Eul}
\end{figure}

\begin{figure}
\centering\includegraphics[width=0.5\textwidth]{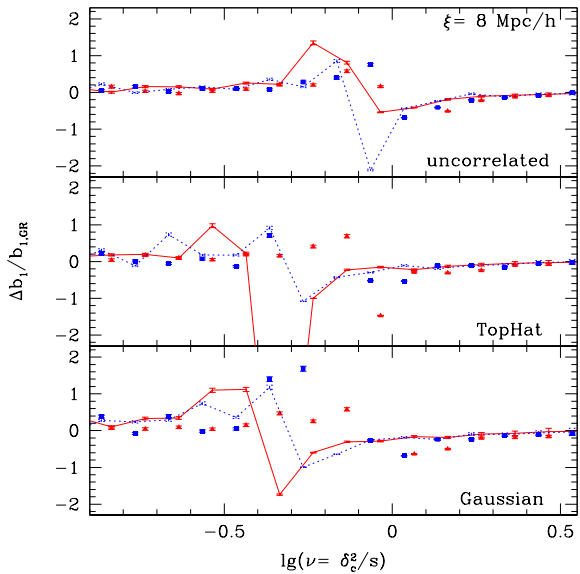}
\caption{(Colour Online) Relative difference in halo bias 
         for chameleon models (Lagrangian environment) to that of GR. 
}
\label{fig:b1Lag}
\end{figure}

Halo bias using more correlated window functions has stronger dependence
on the value of $\delta_0$, which can be seen by comparing the 
spread of bias among the three panels.
Including the chameleon-type fifth force modifies the halo bias and their relative 
changes are shown in figure~\ref{fig:b1Eul} and \ref{fig:b1Lag} for 
the Eulerian and Lagrangian environments respectively.
In both environment definitions the relative difference 
of the halo bias in chameleon models can be of factor of a few, 
regardless of the window functions used. 
Another noticeable signature is the difference of the 
mass range where the halo bias relative difference is most significant: 
when the tophat or Gaussian window filters are used, the 
change in halo bias for most overdense environment (triangles, 
$\delta_0= 0.5$) 
is most significant around $\lg(\nu)=-0.1$ but it is shifted to lower mass
$\lg(\nu)=-0.5$ for underdense environment (solid lines, $\delta_0= -0.5$).
This difference in mass range is getting smaller when $\delta_0$ is less 
extreme (see the difference between squares and dotted lines 
for $\delta_0=\pm 0.35$).
Combining the measurements of
 the halo bias in very overdense and underdense environments to the 
GR predictions (analytical formula or measurements from numerical simulations)
would probably constrain the chameleon models.

\section{Conclusions}

\label{sect:con}

In this paper we investigated the modifications in the mass function and the 
halo 
bias due to the fifth force 
in the framework of excursion set approach.
Two different environment definitions (Eulerian environment 
at $\zeta = 5\ {\rm Mpc}/h$ and Lagrangian environment $\xi = 8 {\rm Mpc}/h$)
are combined with three different smoothing window functions: Gaussian, 
tophat, and sharp-\textit{k}, to study how correlations between 
different scales induced by the window functions interact with the 
fifth force.

The halo formation barrier in the chameleon model depends on the environment 
density $\delta_{\rm env}$ -- in an accompanied paper \citep{ll2012} 
we discussed  the change in the mass function for Eulerian environment 
versus Lagrangian environment for uncorrelated steps 
(i.e. sharp-\textit{k} window filter) using the excursion set approach.
In that case the difference in the two environments only modifies 
the distribution of $\delta_{\rm env}$ and subsequently the 
first crossing distribution. 
Additional effect due to the non-trivial correlations between different 
smoothing scales arises when the smoothing function is Gaussian or tophat.
We used monte-carlo simulations to demonstrate the additional 
correlations indeed modify the first crossing distribution and the
contribution is significant, particularly in intermediate to low mass range.
We then applied the analytical formalism proposed in \citet{ms12} to 
compute the first crossing distribution for correlated steps. 
The analytical prediction matches the monte-carlo simulations well for high 
mass regime. 

Having shown the significance of correlated steps in the unconditional 
first crossing distribution, we examined the conditional mass function 
in chameleon models. 
When the steps in the excursion set are uncorrelated, the 
condition that random walks having passed through some $\delta_0$ at 
a prescribed large scale $S_0$ only alters 
the distribution of $\delta_{\rm env}$ due to the Markovian nature of 
uncorrelated steps.
Correlated steps (the random walk is non-Markovian) 
introduce an additional effect due to the non-trivial 
correlations between $S_0$ and all other scales.
We compared the change in the first crossing distribution due to the 
fifth force for conditional walks to that of unconditional walks and 
found that walks in underdense large scale environment experience
a stronger modification compared to those in overdense large scale
environment --  this effect is observed in  all the three smoothing 
window functions (with different strengths) usually used in the literature.
Hence combining the conditional halo mass function in underdense large
scale environment with the unconditional mass function can potentially 
provide a strong probe for modified gravity.

Finally we investigate how the chameleon models modify 
the halo bias derived from the excursion set approach. 
We found that the fifth force can 
modify the halo bias by a factor of a few 
at intermediate to low mass halos when underdense large scale environment is assumed. 
In addition we found that, when the correlated steps are used, 
the masses at which the halo bias is modified by the chameleon-type fifth force
can be quite different for overdense and underdense environments.
While this difference depends on the value of $\delta_0$, we found that 
 $\Delta\lg(\nu)=0.4$ when $\delta_0$ at 2-$\sigma$ level for
$S_0=0.15^2\delta_{c0}$.

The effect of the correlations between different smoothing scales introduced by 
the use of different smoothing window functions on halo abundance and 
halo bias has been the focus of some the recent studies. 
Its application with the excursion set approach in the chameleon models 
results in further modifications in both halo abundance and the halo bias. 
In particular the change in the conditional mass function and the halo bias
in underdense environment may provide  potentially strong constraints 
on the chameleon models.
Another test we are investigating is the strength of  correlations 
between the halo formation and the surrounding environment by 
measuring halo abundance and clustering under different environments 
in numerical simulations of chameleon models.
This will be left for future study.

\section*{Acknowledgments}

TYL would like to thank Ravi Sheth for discussion. TYL is supported in part by Grant-in-Aid for Young Scientists (22740149) and by WPI Initiative, MEXT, Japan. 
BL is supported by the Royal Astronomical Society and the Department of Physics of Durham University, and acknowledges the host of IPMU where this work was initiated.

%

\label{lastpage}

\end{document}